\documentstyle[amssymb,aps]{revtex}
\begin{document}
\title{Obtaining the Neutrino Mixing Matrix with the Tetrahedral Group}
\author{A. Zee}
\address{Kavli Institute for Theoretical Physics\\
University of California\\
Santa Barbara, CA 93106\\
USA\\
zee@kitp.ucsb.edu\\
(to appear in Physics Letters B)}
\maketitle

\begin{abstract}
We discuss various ``minimalist'' schemes to derive the neutrino mixing
matrix using the tetrahedral group $A_{4}.$
\end{abstract}

\bigskip

{\bf {\large I. Neutrino mixing matrix}}

\medskip The neutrino mixing matrix $V$ relates the neutrino current
eigenstates (denoted by $\nu _{\alpha }$, $\alpha =e,$ $\mu ,$ $\tau ,$ and
coupled by the $W$ bosons to the corresponding charged leptons) to the
neutrino mass eigenstates (denoted by $\nu _{i}$, $i=1,2,3$ and endowed with
definite masses $m_{i})$ according to 
\begin{equation}
\left( 
\begin{array}{l}
\nu _{e} \\ 
\nu _{\mu } \\ 
\nu _{\tau }
\end{array}
\right) =V\left( 
\begin{array}{l}
\nu _{1} \\ 
\nu _{2} \\ 
\nu _{3}
\end{array}
\right)  \label{vdef}
\end{equation}
Thanks to heroic experimental efforts, the neutrino mixing angles have now
been determined \cite{GonGar} to be given by $\sin ^{2}\theta _{12}\sim
0.31,\;\sin ^{2}\theta _{23}\sim 0.50,$ and $\sin ^{2}\theta _{31}\sim 0.01,$
with the mixing angles defined by the standard parametrization (with $%
c_{23}\equiv \cos \theta _{23},$ $s_{23}\equiv \sin \theta _{23},$ and so
forth) 
\begin{eqnarray}
V_{{\rm angular}} &=&V_{23}V_{31}V_{12}  \label{cklike} \\
&=&\left( 
\begin{array}{lll}
1 & 0 & 0 \\ 
0 & c_{23} & s_{23} \\ 
0 & s_{23} & -c_{23}
\end{array}
\right) \left( 
\begin{array}{lll}
c_{31} & 0 & s_{31}e^{-i\phi } \\ 
0 & 1 & 0 \\ 
-s_{31}e^{i\phi } & 0 & c_{31}
\end{array}
\right) \left( 
\begin{array}{lll}
-c_{12} & s_{12} & 0 \\ 
s_{12} & c_{12} & 0 \\ 
0 & 0 & 1
\end{array}
\right) \\
&=&\left( 
\begin{array}{lll}
-c_{31}c_{12} & c_{31}s_{12} & s_{31}e^{-i\phi } \\ 
s_{12}c_{23}+c_{12}s_{23}s_{31}e^{i\phi } & 
c_{12}c_{23}-s_{12}s_{23}s_{31}e^{i\phi } & s_{23}c_{31} \\ 
s_{12}s_{23}-c_{12}c_{23}s_{31}e^{i\phi } & 
c_{12}s_{23}+s_{12}c_{23}s_{31}e^{i\phi } & -c_{23}c_{31}
\end{array}
\right)
\end{eqnarray}
(This parametrization may differ slightly from others in that we take $\det
V_{23}=\det V_{12}=-1$.) The error bars are such that $\theta _{31}$ is
consistent with 0, in which case the $CP$ violating phase $e^{i\phi }$ does
not enter.

We could suppose either that the entries in $V$ represent a bunch of
meaningless numbers possibly varying from village to village in the multiverse landscape as advocated by some theorists of great sophistication or that they point
to some deeper structure or symmetry as some theorists with a more traditional faith in the power of theoretical physics might dare to hope for. It is natural to imagine that there is
a family symmetry \cite{wilzee} linking the three lepton families. Starting
with the standard model we assign (all fermionic fields are left handed) the
lepton doublets $\psi _{a}=\left( 
\begin{array}{l}
\nu _{a} \\ 
l_{a}
\end{array}
\right) $, the lepton singlets $l_{a}^{C}$ ($a=1,2,3),$ and the required
Higgs fields to various representations of a family group$\cite{group}$ $%
G_{F}.$

Indeed, if we guess that $s_{12}=1/\sqrt{3},$ $s_{23}=1/\sqrt{2},$ and $%
s_{31}=0,$ we obtain the attractive mixing matrix 
\begin{equation}
V=\left( 
\begin{array}{rrr}
-{\frac{2}{\sqrt{6}}} & {\frac{1}{\sqrt{3}}} & 0 \\ 
{\frac{1}{\sqrt{6}}} & {\frac{1}{\sqrt{3}}} & {\frac{1}{\sqrt{2}}} \\ 
{\frac{1}{\sqrt{6}}} & {\frac{1}{\sqrt{3}}} & {-\frac{1}{\sqrt{2}}}
\end{array}
\right) .  \label{theV}
\end{equation}
first proposed by Harrison, Perkins and Scott\cite{hps}. Later, X. G. He and
I independently arrived at the same Ansatz\cite{hz1} Also, this mixing
matrix (but curiously, with the first and second column interchanged) was
first suggested by Wolfenstein more than 20 years ago\cite{wolf78} based on
some considerations involving the permutation group $S_{3}$. It has
subsequently been studied extensively by Harrison, Perkins and Scott\cite{21}
, and by Xing\cite{22}. Attempts to derive this mixing matrix have been
discussed by Low and Volkas\cite{lv}\cite{low}. A parametrization of the
experimental data in terms of deviation from $V$ is given in \cite{zeepara}.
Following Wolfenstein and defining $\nu _{x}\equiv (\nu _{\mu }+\nu _{\tau
})/\sqrt{2}$ and $\nu _{y}\equiv (\nu _{\mu }-\nu _{\tau })/\sqrt{2},$ we
see that (\ref{theV}) says that the mass eigenstates are given by 
\begin{equation}
\nu _{1}=-\sqrt{\frac{2}{3}}\nu _{e}+{\frac{1}{\sqrt{3}}}\nu _{x},\nu _{2}={%
\frac{1}{\sqrt{3}}}\nu _{e}+\sqrt{\frac{2}{3}}\nu _{x},\nu _{3}=\nu _{y}
\end{equation}
The basis $\{\nu _{1},\nu _{2}\}$ is rotated from $\{\nu _{e},\nu _{x}\}$
through $\arcsin (1/\sqrt{3})\sim 35^{o}.$

In this paper we will take the neutrinos to be Majorana\cite{kay} as seems
likely, so that we have in the Lagrangian the mass term ${\cal L}=-\nu
_{\alpha }M_{\alpha \beta }C\nu _{\beta }+h.c.$ where $C$ denotes the charge
conjugation matrix. Thus, the neutrino mass matrix $M$ is symmetric. Also,
for the sake of simplicity we will assume $CP$ conservation so that $M$ is
real. With this simplification, the orthogonal transformation $V^{T}MV$
produces a diagonal matrix with diagonal elements $m_{1},m_{2}$ and $m_{3}.$
We are free to multiply $V$ on the right by some diagonal matrix whose
diagonal entries are equal to $\pm 1$. This merely multiplies each of the
columns in $V$ by an arbitrary sign. Various possible phases have been
discussed in detail in the literature\cite{nieves,sign}.

At present, we have no understanding of the neutrino masses just as we have
no understanding of the charged lepton and quark masses. The well-known
solar and atmospheric neutrino experiments have determined respectively that 
$\Delta m_{\odot }^{2}=m_{2}^{2}-m_{1}^{2}\sim 8\times 10^{-5}$ $eV^{2}$ and 
$\Delta m_{{\rm atm}}^{2}=m_{3}^{2}-m_{2}^{2}\sim \pm 2.4\times 10^{-3}$ $%
eV^{2}.$ The sign of $\Delta m_{{\rm atm}}^{2}$ is currently unknown, while $%
\Delta m_{\odot }^{2}\ $has to be positive in order for the
Mikheyev-Smirnov-Wolfenstein resonance to occur inside the sun. We could
have either the so-called normal hierarchy in which $|m_{3}|>|m_{2}|\sim
|m_{1}|$ or the inverted hierarchy $|m_{3}|<|m_{2}|\sim |m_{1}|.$

\bigskip

{\bf {\large II. Family symmetry and the tetrahedral group}}

\smallskip

For some years, E. Ma \cite{ma1} has advocated choosing the discrete group $%
A_{4},$ namely the symmetry group of the tetrahedron, as $G_{F}.$ With
various collaborators he has written a number of interesting papers \cite
{ma2}\cite{ma3}\cite{ma4}\cite{ma5} using $A_{4}$ to study the lepton sector.

For the convenience of the reader and to set the notation, we give a concise
review of the relevant group theory. Evidently, $A_{4}$ is a subgroup of $%
SO(3)$ (which was often used in the early literature on family symmetry but
which has proved to be too restrictive.) Since the tetrahedron lives in
3-dimensional space, $A_{4}$ has a natural 3-dimensional representation
denoted by $\underline{3}$ suggestive of the 3 families observed in nature.
The tetrahedron has 4 vertices and thus $A_{4}$ is also formed by the even
permutations of 4 objects so that $A_{4}$ has $4!/2=12$ elements which could
be represented as elements of $SO(3).$ Besides the identity $I=\left( 
\begin{array}{lll}
1 & 0 & 0 \\ 
0 & 1 & 0 \\ 
0 & 0 & 1
\end{array}
\right) ,$we have the 3 rotations through 180$^{o}$ $r_{1}=\left( 
\begin{array}{lll}
1 & 0 & 0 \\ 
0 & -1 & 0 \\ 
0 & 0 & -1
\end{array}
\right) ,r_{2}=\left( 
\begin{array}{lll}
-1 & 0 & 0 \\ 
0 & 1 & 0 \\ 
0 & 0 & -1
\end{array}
\right) ,$ and $r_{3}=\left( 
\begin{array}{lll}
-1 & 0 & 0 \\ 
0 & -1 & 0 \\ 
0 & 0 & 1
\end{array}
\right) .$ Then we have the cyclic permutation $c=\left( 
\begin{array}{lll}
0 & 0 & 1 \\ 
1 & 0 & 0 \\ 
0 & 1 & 0
\end{array}
\right) ,$ which together with $r_{1}cr_{1},r_{2}cr_{2}$ and $r_{3}cr_{3},$
form an equivalence class with 4 members. Finally we have the anticyclic
permutation $a=\left( 
\begin{array}{lll}
0 & 1 & 0 \\ 
0 & 0 & 1 \\ 
1 & 0 & 0
\end{array}
\right) ,$ which together with $r_{1}ar_{1},r_{2}ar_{2}$ and $r_{3}ar_{3},$
form another equivalence class with 4 members. Thus, the 12 elements belong
to 4 equivalence classes with membership 1, 3, 4, and 4, which tells us that
there are 4 irreducible representations with dimension $d_{j}$ such that $%
\sum_{j}d_{j}^{2}=12$ which has the unique solution $d_{1}=d_{2}=d_{3}=1$
and $d_{4}=3.$ The natural 3-dimensional representation $\underline{3}$ has
just been displayed explicitly.

The multiplication of representations is easy to work out by using the
following trick. Start with the familiar multiplication within $SO(3):$ $%
\underline{3}\times \underline{3}=\underline{1}+\underline{3}+\underline{5}.$
\bigskip Given two vectors $\vec{x}$ and $\vec{y}$ of $SO(3),$ the $%
\underline{3}$ is of course given by the cross product $\vec{x}\times $ $%
\vec{y}$ while the $\underline{5}$ is composed of the symmetric combinations 
$x_{2}y_{3}+x_{3}y_{2},$ $x_{3}y_{1}+x_{1}y_{3},$ $x_{1}y_{2}+x_{2}y_{1},$
together with the 2 diagonal traceless combinations $%
2x_{1}y_{1}-x_{2}y_{2}-x_{3}y_{3}$ and $x_{2}y_{2}-x_{3}y_{3}$. Upon
restriction of $SO(3)$ to $A_{4}$ the $\underline{5}$ evidently decompose
into $\underline{5}\rightarrow \underline{3}+\underline{1}^{\prime }+%
\underline{1}^{\prime \prime }$ with the $\underline{3}$ given by the 3
symmetric combinations just displayed. The $\underline{1}^{\prime }$ and $%
\underline{1}^{\prime \prime }$ could be taken respectively as linear
combinations of the 2 traceless combinations just given: 
\begin{equation}
\underline{1}^{\prime }\sim u^{\prime }=x_{1}y_{1}+\omega x_{2}y_{2}+\omega
^{2}x_{3}y_{3}
\end{equation}
and 
\begin{equation}
\underline{1}^{\prime \prime }\sim u^{\prime \prime }=x_{1}y_{1}+\omega
^{2}x_{2}y_{2}+\omega x_{3}y_{3}
\end{equation}
with $\omega \equiv e^{i2\pi /3}$ the cube root of unity so that 
\begin{equation}
1+\omega +\omega ^{2}=0.
\end{equation}
It is perhaps worth emphasizing the obvious, that while $\underline{1}%
^{\prime }$ and $\underline{1}^{\prime \prime }$ furnish 1-dimensional
representations of $A_{4}$ they are not invariant under $A_{4}.$ For
example, under the cyclic permutation $c,$ $u^{\prime }\rightarrow \omega
u^{\prime }$ and $u^{\prime \prime }\rightarrow \omega ^{2}u^{\prime \prime
}.$ Evidently $\underline{1}^{\prime }$ $\times $ $\underline{1}^{\prime
\prime }=\underline{1},$ $\underline{1}^{\prime }$ $\times $ $\underline{1}
^{\prime }=\underline{1}^{\prime \prime },$and $\underline{1}^{\prime \prime
}$ $\times $ $\underline{1}^{\prime \prime }=\underline{1}^{\prime },$ and
also ($\underline{1}^{\prime })^{*}=\underline{1}^{\prime \prime }.$

Thus, under $A_{4}$ we have $\underline{3}\times \underline{3}=\underline{1}+%
\underline{1}^{\prime }+\underline{1}^{\prime \prime }+\underline{3}+%
\underline{3}.$ It is perhaps also worth remarking that the two $\underline{3%
}$'s on the right hand side may be taken as $%
(x_{2}y_{3},x_{3}y_{1},x_{1}y_{2})$ and $(x_{3}y_{2},x_{1}y_{3},x_{2}y_{1}).$
The existence of 3 inequivalent 1-dimensional representations also suggests
the relevance of $A_{4}$ to the family problem. I can't resist mentioning here the possibly physically irrelevant fact that \cite{car} alone among all the alternating groups $A_{n}$'s the group $A_{4}$ is not simple.

\bigskip

{\bf {\large III. A minimalist framework}}

Given these attractive features of $A_{4},$ there has been, perhaps not
surprisingly, a number of recent attempts \cite{ma1}\cite{alta}\cite{babuhe}
to derive $V$ using $A_{4}.$ In our opinion, they all appear to involve a
rather elaborate framework, for example supersymmetry, higher dimensional
spacetime, and so on. Within this recent literature Ma \cite{malatest} has
produced a particularly interesting and relatively economical scheme in
which the neutrino mixing matrix depends on a parameter such that when that
parameter takes on ``reasonable'' values the matrix $V$ as given in (\ref
{theV}) is recovered approximately.

The guiding philosophy of this paper is that we would like to have as
minimal a theoretical framework as possible.

Within a minimalist framework, charged lepton masses are generated by the
dimension 4 operator 
\begin{equation}
O_{4}=\varphi ^{\dagger }l^{C}\psi
\end{equation}
Here $\varphi $ denotes generically the standard Higgs doublet, of which we
may have more than one. According to a general low energy effective field
theory analysis \cite{wein}\cite{wz}\cite{welzee} neutrino masses are
generated by the dimension 5 operator 
\begin{equation}
O_{5}=(\xi \tau _{2}\psi )C(\xi ^{\prime }\tau _{2}\psi )  \label{dim5}
\end{equation}
in the Lagrangian. Here $\xi $ and $\xi ^{\prime }$ denote various Higgs
doublets that may or may not be the same as the $\varphi $'s. We will
suppress the charge conjugation matrix $C$ and the Pauli matrix $\tau _{2}$
in what follows. It is important to emphasize that the analysis leading up
to (\ref{dim5}) is completely general and depends only on $SU(2)\times U(1),$
and not on which dynamical model you believe in, be it the seesaw mechanism
or some other mechanism (such as the model in \cite{zeemodel}).

We suppose that the family symmetry remains unbroken down to the scale of $%
SU(2)\times U(1)$ breaking, so that the operators $O_{4}$ and $O_{5}$ have
to be singlets under $G_{F}$. As is completely standard, when $\varphi ,$ $%
\xi ,$ and $\xi ^{\prime }$ acquire vacuum expectation values, $SU(2)\times
U(1)$ and $G_{F}$ are broken and the neutrinos acquire masses given by the
mass matrix $M_{\nu }\propto <\xi ><\xi ^{^{\prime }}>$ as well as the
charged leptons. (Henceforth, for a Higgs doublet $\xi $ we use $<\xi >$ to
denote the vacuum expectation of the lower electrically neutral component of 
$\xi .)$

Let $M_{\nu }$ be diagonalized by $U_{\nu }^{T}M_{\nu }U_{\nu }=D_{\nu }$ so
that the 3 neutrino fields that appear in $\psi _{a}$ are related to the
neutrino fields $\nu ^{m}$ with definite masses by $\nu =U_{\nu }\nu ^{m}.$
Similarly, let the 3 charged left handed lepton fields $l$ that appear in $%
\psi _{a}$ be related to the physical charged lepton fields $l^{m}$ by $%
l=U_{l}l^{m}.$ Then $\psi _{a}=\left( 
\begin{array}{l}
(U_{\nu })_{ab}\nu _{b}^{m} \\ 
(U_{l})_{ab}l_{b}^{m}
\end{array}
\right) $so that the neutrino mixing matrix as defined in (\ref{vdef}) is
given by $V=U_{l}^{\dagger }U_{\nu }.$ One difficulty in constructing a
theory for $V$ is that it arises from the ``mismatch'' between two rotations 
$U_{l}$ and $U_{\nu }.$

As in turns out, in our model building efforts, we often have to forbid the $%
\varphi $'s that appear in $O_{4}$ from appearing in $O_{5}.$ This could
easily be implemented by imposing a discrete symmetry under which $\varphi
\rightarrow e^{i\varkappa }\varphi ,$ $l^{C}\rightarrow e^{i\varkappa }l^{C}$
(where $e^{i\varkappa }\neq -1$ is some appropriate phase factor), with all
other fields unaffected. We will leave this implicit in what follows.

Within the minimalist framework outlined here we offer some possible
schemes. None of these could be said to be terribly compelling but at least
we keep within the usual rules of the model building literature. The various
schemes, depending on what representations of $A_{4}$ we choose for the
various fields $\psi ,$ $l^{C}$, and $\varphi $, could be listed
systematically.

{\bf {\large IV. Model A}}

\bigskip We first try the assignment $\psi \sim \underline{3},$ $l^{C}\sim 
\underline{1},$ $\underline{1}^{\prime }$, and $\underline{1}^{\prime \prime
},$ and $\varphi \sim \underline{3}.$ The Lagrangian then contains the terms

\begin{eqnarray}
&&h_{1}l_{1}^{C}(\varphi _{1}^{\dagger }\psi _{1}+\varphi _{2}^{\dagger
}\psi _{2}+\varphi _{3}^{\dagger }\psi _{3})+h_{2}l_{2}^{C}(\omega \varphi
_{1}^{\dagger }\psi _{1}+\varphi _{2}^{\dagger }\psi _{2}+\omega ^{2}\varphi
_{3}^{\dagger }\psi _{3})+h_{3}l_{3}^{C}(\omega ^{2}\varphi _{1}^{\dagger
}\psi _{1}+\varphi _{2}^{\dagger }\psi _{2}+\omega \varphi _{3}^{\dagger
}\psi _{3})  \label{modelAchargedlep} \\
&=&\left( 
\begin{array}{lll}
l_{1}^{C} & l_{2}^{C} & l_{3}^{C}
\end{array}
\right) \left( 
\begin{array}{lll}
h_{1} & 0 & 0 \\ 
0 & h_{2} & 0 \\ 
0 & 0 & h_{3}
\end{array}
\right) \left( 
\begin{array}{lll}
1 & 1 & 1 \\ 
\omega & 1 & \omega ^{2} \\ 
\omega ^{2} & 1 & \omega
\end{array}
\right) \left( 
\begin{array}{lll}
\varphi _{1}^{\dagger } & 0 & 0 \\ 
0 & \varphi _{2}^{\dagger } & 0 \\ 
0 & 0 & \varphi _{3}^{\dagger }
\end{array}
\right) \left( 
\begin{array}{l}
\psi _{1} \\ 
\psi _{2} \\ 
\psi _{3}
\end{array}
\right)
\end{eqnarray}
It is natural for the 3 $<$ $\varphi _{\alpha }>=v_{\alpha }$'s to be equal
since $A_{4}$ requires that the coefficients of $\varphi _{\alpha }^{\dagger
}\varphi _{\alpha }$ and of ($\varphi _{\alpha }^{\dagger }\varphi _{\alpha
})^{2}$ in the potential be independent of $\alpha =1,2,3$. (See the
Appendix for a more detailed analysis.) If so, then upon spontaneous gauge
symmetry breaking we obtain 
\begin{equation}
\left( 
\begin{array}{lll}
l_{1}^{C} & l_{2}^{C} & l_{3}^{C}
\end{array}
\right) \left( 
\begin{array}{lll}
m_{e} & 0 & 0 \\ 
0 & m_{\mu } & 0 \\ 
0 & 0 & m_{\tau }
\end{array}
\right) \left( 
\begin{array}{lll}
1 & 1 & 1 \\ 
\omega & 1 & \omega ^{2} \\ 
\omega ^{2} & 1 & \omega
\end{array}
\right) \left( 
\begin{array}{l}
\psi _{1} \\ 
\psi _{2} \\ 
\psi _{3}
\end{array}
\right)
\end{equation}
with $m_{e}=h_{1}v$ and so on. It is useful to define the ``magic'' matrix%
\cite{4thpower} 
\begin{equation}
A=\left( 
\begin{array}{lll}
1 & 1 & 1 \\ 
\omega & 1 & \omega ^{2} \\ 
\omega ^{2} & 1 & \omega
\end{array}
\right)
\end{equation}
Then $l^{m}=\frac{1}{\sqrt{3}}Al$ or $l=\sqrt{3}A^{-1}l^{m}$ so that $%
U_{l}^{\dagger }=(\sqrt{3}A^{-1})^{\dagger }=\frac{1}{\sqrt{3}}A$

The crucial observation at this point is that the sum of the first and third
columns in $A$ gives $\left( 
\begin{array}{l}
2 \\ 
-1 \\ 
-1
\end{array}
\right) $ and that the difference of the first and third columns in $A$
gives $\sqrt{3}i\left( 
\begin{array}{l}
0 \\ 
1 \\ 
-1
\end{array}
\right) ,$ which up to some overall factors are precisely the first and
third column respectively in the desired $V$ in (\ref{theV}). In other
words, if 
\begin{equation}
U_{\nu }=\frac{1}{\sqrt{2}}\left( 
\begin{array}{lll}
1 & 0 & -1 \\ 
0 & \sqrt{2} & 0 \\ 
1 & 0 & 1
\end{array}
\right) ,  \label{unu}
\end{equation}
then $U_{l}^{\dagger }U_{\nu }=V\Phi $ with $V$ the desired mixing matrix in
(\ref{theV}) and the diagonal phase matrix $\Phi $ with the diagonal
elements $-1,1,$ and $-i.$ Thus, if we could obtain $U_{\nu }$ we would
achieve our goal of deriving $V.$

We recognize that $U_{\nu }$ is just a rotation through 45$^{o}$ in the $%
(1-3)$ plane. Recalling that $U_{\nu }$ is determined by requiring $U_{\nu
}^{T}M_{\nu }U_{\nu }=D_{\nu }$ be diagonal we see that if we could obtain
an $M_{\nu }$ of the form 
\begin{equation}
M_{\nu }=\left( 
\begin{array}{lll}
\alpha & 0 & \beta \\ 
0 & \gamma & 0 \\ 
\beta & 0 & \alpha
\end{array}
\right)  \label{mnuform}
\end{equation}
(note that the $2\times 2$ matrix in the $(1-3)$ sector has equal diagonal
elements) then we are done. Our discussion here overlaps with that given
recently by Babu and He\cite{babuhe}; however, their discussion is given in
the context of a much more elaborate scheme involving supersymmetry.

Referring to (\ref{dim5}) we see that by imposing a discrete symmetry $K_{2}$
under which $\psi _{2}\rightarrow -\psi _{2},$ $\varphi _{2}\rightarrow
-\varphi _{2},$ with all other fields unaffected, or equivalently a discrete
symmetry $K_{13}$ under which $\psi _{1}\rightarrow -\psi _{1},\psi
_{3}\rightarrow -\psi _{3},$ $\varphi _{1}\rightarrow -\varphi _{1},$ $%
\varphi _{3}\rightarrow -\varphi _{3},$ with all other fields unaffected, we
can obtain the texture zeroes in (\ref{mnuform}), but unfortunately this
does not imply that $($ $M_{\nu })_{11}=($ $M_{\nu })_{33}$. Furthermore, $%
K_{13}$ is just the element $r_{2}$ of $A_{4}$ and so it does not commute
with $A_{4}.$ Note that upon the $\varphi $'s acquiring equal vacuum
expectation values, $A_{4}$ is broken down to a $Z_{3}$ generated by $%
\{I,c,a\}$ and unfortunately $r_{2}$ does not belong in $Z_{3}.$ Perhaps
there is a more attractive scheme in which a reflection symmetry like $K_{2}$
could emerge effectively.

In another attempt to obtain an $M_{\nu }$ of the form in (\ref{mnuform}) we
introduce Higgs doublets $\chi $ and $\xi $ transforming as $\underline{1}$
and $\underline{3}$ respectively. We then have three types of $O_{5}$
operators, namely $(\chi \psi )^{2},$ $(\chi \psi )(\xi \psi ),$ and $(\xi
\psi )^{2}.$ As mentioned earlier, we impose a discrete symmetry to forbid $%
\varphi $ from participating in $O_{5}.$ As discussed in the Appendix, we
could naturally suppose that the vacuum expectation value of $\xi $ points
in the 2-direction, that is, $<\xi _{2}>\neq 0$ with $<\xi _{1}>=<\xi
_{3}>=0.$ Let us now list how the different $O_{5}$ operators contribute to $%
M_{\nu }$ upon $\chi $ and $\xi _{2}$ acquiring a vacuum expectation value.
The operator $(\chi \psi )^{2}$ contributes a term proportional to the
identity matrix. Next, $(\chi \psi )(\xi \psi ),$ which is formed by $%
\underline{3}\times \underline{3}\times \underline{3},$ consists of two
terms, corresponding to the two ways of obtaining a $\underline{3}$ upon
multiplying $\underline{3}\times \underline{3}.$ One term has the form $\chi
(\psi _{1}\xi _{2}\psi _{3}+\psi _{2}\xi _{3}\psi _{1}+\psi _{3}\xi _{1}\psi
_{2}),$ with the other term having an analogous form. Thus, the operator $%
(\chi \psi )(\xi \psi )$ contributes the term denoted by $\beta $ in (\ref
{mnuform}). Finally, the operator $(\xi \psi )^{2}$ actually denotes
schematically 4 different operators since it is formed by $(\underline{3}%
\times \underline{3})\times (\underline{3}\times \underline{3})$ and this
contains $\underline{1}\times \underline{1},\underline{1}^{\prime }\times 
\underline{1}^{\prime \prime },$ $\underline{3}\times \underline{3},$ $\underline{3}\times \underline{3},$ and $%
\underline{3}\times \underline{3}$, corresponding respectively to the
operators 
\begin{equation}
(\xi _{1}\psi _{1}+\xi _{2}\psi _{2}+\xi _{3}\psi _{3})^{2},  \label{O5a}
\end{equation}
\begin{equation}
(\xi _{1}\psi _{1}+\omega \xi _{2}\psi _{2}+\omega ^{2}\xi _{3}\psi
_{3})(\xi _{1}\psi _{1}+\omega ^{2}\xi _{2}\psi _{2}+\omega \xi _{3}\psi
_{3}),  \label{O5b}
\end{equation}
\begin{equation}
(\xi _{2}\psi _{3},\xi _{3}\psi _{1},\xi _{1}\psi _{2}).(\xi _{3}\psi
_{2},\xi _{1}\psi _{3},\xi _{2}\psi _{1})=\xi _{1}\psi _{2}\xi _{2}\psi
_{1}+\xi _{2}\psi _{3}\xi _{3}\psi _{2}+\xi _{3}\psi _{1}\xi _{1}\psi _{3},
\label{O5c}
\end{equation}
\begin{equation}
(\xi _{3}\psi _{2},\xi _{1}\psi _{3},\xi _{2}\psi _{1}).(\xi _{3}\psi
_{2},\xi _{1}\psi _{3},\xi _{2}\psi _{1})=(\xi _{3}\psi _{2})^{2}+(\xi
_{1}\psi _{3})^{2}+(\xi _{2}\psi _{1})^{2}  \label{O5d}
\end{equation}
and 
\begin{equation}
(\xi _{2}\psi _{3},\xi _{3}\psi _{1},\xi _{1}\psi _{2}).(\xi _{2}\psi
_{3},\xi _{3}\psi _{1},\xi _{1}\psi _{2})=(\xi _{1}\psi _{2})^{2}+(\xi
_{2}\psi _{3})^{2}+(\xi _{3}\psi _{1})^{2}.  \label{O5e}
\end{equation}
(This is essentially the same as the analysis of an $A_{4}$ invariant Higgs
potential given in the Appendix.) Upon $\xi _{2}$ acquiring a vacuum
expectation value, we obtain respectively $\psi _{2}\psi _{2},$ $\psi
_{2}\psi _{2},$ $0,$ $\psi _{1}\psi _{1}$ and $\psi _{3}\psi _{3}.$
Unfortunately, the effective coupling constants in front of the operator in (%
\ref{O5d}) and (\ref{O5e}) are in general not equal to each other and thus
we obtain an $M_{\nu }$ of the form 
\begin{equation}
M_{\nu }=\left( 
\begin{array}{lll}
\alpha -\varepsilon  & 0 & \beta  \\ 
0 & \gamma  & 0 \\ 
\beta  & 0 & \alpha +\varepsilon 
\end{array}
\right) 
\end{equation}
rather than the $M_{\nu }$ in (\ref{mnuform}). To set $\varepsilon $ to 0 we
would have impose a discrete interchange symmetry $P_{13}$ which
interchanges the indices 1 and 3 but unfortunately, just as before for $%
K_{13}$, $P_{13}$ does not commute with $A_{4}.$

At this point, we could only suppose that $\varepsilon $ is small compared
to $\beta ,$ in which case $U_{\nu }$ is perturbed from the desired $U_{\nu }
$ in (\ref{unu}) to 
\begin{equation}
U_{\nu }=\frac{1}{\sqrt{2}}\left( 
\begin{array}{lll}
1 & 0 & -1 \\ 
0 & \sqrt{2} & 0 \\ 
1 & 0 & 1
\end{array}
\right) \left( 
\begin{array}{lll}
1 & 0 & -\frac{\varepsilon }{2\beta } \\ 
0 & 1 & 0 \\ 
\frac{\varepsilon }{2\beta } & 0 & 1
\end{array}
\right) 
\end{equation}
The resulting deviation from the $V$ in (\ref{theV}) may be interesting
phenomenologically. In particular, $V_{e3}\simeq -\frac{\varepsilon }{\sqrt{6%
}\beta }$ is no longer identically 0. In \cite{zeepara} it was advocated
that experimental data be parametrized as a deviation from $V$ in (\ref{theV}%
) as discussed in section III there.

In this scheme, the neutrino masses come out to be $\alpha - \sqrt{\beta^{2}+\varepsilon ^{2}} $ ,$\gamma ,$
and $\alpha +\sqrt{\beta^{2}+\varepsilon ^{2}} $ and thus both the normal hierarchy and the inverse
hierarchy could be accomodated by suitable tuning, but there is no true
understanding of neutrino masses as remarked earlier.

\bigskip {\bf {\large V. Model B}}

Following Ma\cite{malatest}, we take $\psi \sim \underline{3},$ $l^{C}\sim 
\underline{3}$, and $\varphi \sim \underline{1},\underline{1}^{\prime },$
and $\underline{1}^{\prime \prime }.$ In other words, we have 3 Higgs
doublets $\varphi $ each transforming as a singlet under $A_{4}.$ The
Lagrangian then contains the terms 
\begin{equation}
h_{1}\varphi _{1}^{\dagger }(l_{1}^{C}\psi _{1}+l_{2}^{C}\psi _{2}+\psi
_{3}l_{3}^{C}\psi _{3})+h_{2}\varphi _{2}^{\dagger }(l_{1}^{C}\psi
_{1}+\omega ^{2}l_{2}^{C}\psi _{2}+\omega \psi _{3}l_{3}^{C}\psi
_{3})+h_{3}\varphi _{3}^{\dagger }(l_{1}^{C}\psi _{1}+\omega l_{2}^{C}\psi
_{2}+\omega ^{2}\psi _{3}l_{3}^{C}\psi _{3})
\end{equation}
Upon the $\varphi $'s acquiring vacuum expectation values $v$ we obtain a
diagonal charged lepton mass matrix, with the charged lepton masses given by
the absolute values of $h_{1}v_{1}+$ $h_{2}v_{2}+h_{3}v_{3},$ $h_{1}v_{1}+$ $%
\omega ^{2}h_{2}v_{2}+\omega h_{3}v_{3},$ and $h_{1}v_{1}+$ $\omega
h_{2}v_{2}+\omega ^{2}h_{3}v_{3}.$ All that matters here for our purposes is
that we have enough freedom to match the observed masses $m_{e},$ $m_{\mu },$
and $m_{\tau }.$ The salient point here is that $U_{l}=I$, so that we only
have to worry about getting the desired $U_{\nu }.$

As is obvious and as was discussed in \cite{zeepara} and in \cite{hz1}, in a
basis in which the charged lepton mass matrix is already diagonal, the
neutrino mass matrix $M_{\nu }$ is of course determined in terms of the
three neutrino masses and the neutrino mixing matrix $V.$ Call the three
column vectors in the mixing matrix $\vec{v}_{i}$. Then $M_{\nu }$ is given
by 
\begin{equation}
M_{\nu }=\sum_{i=1}^{3}m_{i}\vec{v}_{i}(\vec{v}_{i})^{T}.
\end{equation}
In particular, if we believe in the $V$ in (\ref{theV}) we have 
\begin{equation}
M_{\nu }=\frac{m_{1}}{6}\left( 
\begin{array}{lll}
4 & -2 & -2 \\ 
-2 & 1 & 1 \\ 
-2 & 1 & 1
\end{array}
\right) +\frac{m_{2}}{3}\left( 
\begin{array}{lll}
1 & 1 & 1 \\ 
1 & 1 & 1 \\ 
1 & 1 & 1
\end{array}
\right) +\frac{m_{3}}{2}\left( 
\begin{array}{lll}
0 & 0 & 0 \\ 
0 & 1 & -1 \\ 
0 & -1 & 1
\end{array}
\right)  \label{massmatrix}
\end{equation}

With $A_{4}$ it is natural to obtain the matrix $M_{D}\equiv \left( 
\begin{array}{lll}
1 & 1 & 1 \\ 
1 & 1 & 1 \\ 
1 & 1 & 1
\end{array}
\right) $ and the identity matrix. In particular, if we introduce a Higgs
doublets $\xi $ transforming as $\underline{3}$ under $A_{4}$ and arrange
the Higgs potential such that the 3 vacuum expectation values $<$ $\xi
_{1}>=<$ $\xi _{2}>=<$ $\xi _{3}>$ are equal. We then see from the list of
operators of the form $(\xi \varphi )(\xi \varphi )$ given in (\ref{O5a}-\ref
{O5d}) at the end of the last section that we obtain for $M_{\nu }$ an
arbitrary linear combination of $M_{D}$ and the identity matrix, which is
not what we want.

In \cite{hz1}, in discussing the neutrino mass matrix, we proposed a basis
of 3 matrices other than those that appear in (\ref{massmatrix}). First, the
3 column vectors in $V$ are the eigenvectors of the matrix 
\begin{equation}
M_{0}=a\left( 
\begin{array}{lll}
2 & 0 & 0 \\ 
0 & -1 & 3 \\ 
0 & 3 & -1
\end{array}
\right)  \label{m0}
\end{equation}
with eigenvalues $m_{1}=m_{2}=2a,$ and $m_{3}=-4a.$ (The parameter $a$
merely sets the overall scale.) Thus, with $M_{0}$ as the mass matrix $%
\Delta m_{21}^{2}=0$ and this pattern reproduces the data $|\Delta
m_{21}^{2}|/|\Delta m_{32}^{2}|\ll 1$ to first approximation. Because of the
degeneracy in the eigenvalue spectrum, $V$ is not uniquely determined. To
determine $V,$ and at the same time to split the degeneracy between $m_{1}\ $
and $m_{2},$ we perturb $M_{0}$ to $M=M_{0}+\varepsilon aM_{D}.$ The matrix $%
M_{D}$ is evidently a projection matrix that projects the first and third
columns in $V$ to zero. Thus, the eigenvalues are given by $%
m_{1}=2a,m_{2}=2a(1+3\varepsilon /2),$ and $m_{3}=-4a,$ where to the lowest
order $\varepsilon =\Delta m_{21}^{2}/\Delta m_{32}^{2}$ and $a^{2}=\Delta
m_{32}^{2}/12$. Finally, to break the relation $|m_{3}|=2|m_{1}|\simeq
2|m_{2}|$ we can always add to $M$ a term proportional to the identity
matrix. But it seems difficult to get the matrix in (\ref{m0}) using $A_{4}$
alone.

\bigskip {\bf {\large VII. Other possibilities and conclusion}}

Given that $A_{4}$ has only 4 distinct representations we could of course
systematically go through all possibilities. Thus, next we could take $\varphi
\sim \underline{3},$ $l^{C}\sim \underline{3}$, and $\psi \sim \underline{1},%
\underline{1}^{\prime },$ and $\underline{1}^{\prime \prime }.$ The charged
lepton mass term would have a form analogous to that given in (\ref
{modelAchargedlep}). But clearly, if we now assume the $<\varphi _{\alpha }>$%
's to be equal, we once again get the matrix $A$ but now acting on $l^{C}$
instead of on $\psi .$ Note that if we assign $\psi _{2}\sim \underline{1}$
and $\psi _{1},$ $\psi _{3}$ to $\underline{1}^{\prime }$ and $\underline{1}%
^{\prime \prime }$ respectively and introduce a Higgs doublets $\chi $
transforming as $\underline{1},$ we get via the operator $O_{5}$ a neutrino
mass matrix $M_{\nu }$ of the form in (\ref{mnuform}) but with $\alpha =0.$

Another possibility is to assign $\psi ,\varphi ,$ and $l^{C}$ all to the $%
\underline{3}$ in which case the charged lepton mass matrix is generated by
two terms, $h(\varphi _{1}^{\dagger }l_{2}^{C}\psi _{3}+\varphi
_{2}^{\dagger }l_{3}^{C}\psi _{1}+\varphi _{3}^{\dagger }l_{1}^{C}\psi _{2})$
and $h^{\prime }(\varphi _{1}^{\dagger }l_{3}^{C}\psi _{2}+\varphi
_{2}^{\dagger }l_{1}^{C}\psi _{3}+\varphi _{3}^{\dagger }l_{2}^{C}\psi
_{1}). $ If we assume the $<\varphi _{\alpha }>$'s to be equal, then the
three charged lepton masses are given in terms of only two parameters.

In conclusion, we have discusssed various schemes to obtain a particularly
attractive neutrino mixing matrix that closely approximates the data.
Instead of detailed models, we use a low energy effective field theory
approach, allowing only Higgs doublets to survive down to the electroweak
scale. We have also explicitly made the restrictive assumption that $A_{4}$
survives down to the $SU(2)\times U(1)$ breaking scale. Of course, if Higgs
triplets could also be used, as for example in \cite{ma2}, or if $A_{4}$ is
broken at higher scale (for example by the coupling of the $\varphi $'s to
the singlet scalar field $h$ in the model in \cite{zeemodel}), then many
more possibilities open up and one could go beyond the discussion given
here. We have been intentionally restrictive here. Ultimately, of course,
any disucssion of neutrino mixing shuld be given in a grand unified
framework (for recent attempts, see for example \cite{ross}\cite{king} in
which neutrino masses, as well as quark masses and mixing, are also
``explained.'' We do not attempt this more ambitious program in this paper.

\bigskip

{\bf {\large Appendix}}

\bigskip

We need to study the Higgs potential for several $SU(2)\times U(1)$ Higgs
doublet $\varphi $'s which transform according to various representations
under $A_{4}.$ For the sake of simplicity, here we restrict ourselves to the
Higgs potential for a single $SU(2)\times U(1)$ Higgs doublet $\varphi $
which transform like a $\underline{3}$ under $A_{4}.$ Hopefully, the
conclusions reached with this restricted analysis continue to hold when the
couplings between different Higgs doublets are small. The multiplication $%
\underline{3}\times \underline{3}=\underline{1}+\underline{1}^{\prime }+%
\underline{1}^{\prime \prime }+\underline{3}+\underline{3}$ tells us that
there is only one quadratic invariant $s=$ $\varphi _{1}^{\dagger }\varphi
_{1}+\varphi _{2}^{\dagger }\varphi _{2}+\varphi _{3}^{\dagger }\varphi
_{3}. $

Since ( $\underline{3}\times \underline{3})\times (\underline{3}\times 
\underline{3})$ contains $\underline{1}$ four times, corresponding to $%
\underline{1}\times \underline{1},$ $\underline{1}^{\prime }\times 
\underline{1}^{\prime \prime },$ $\underline{3}\times \underline{3},$ $\underline{3}\times \underline{3},$ and $%
\underline{3}\times \underline{3},$ we should have 5 quartic invariants. The
obvious quartic invariant is $q=s^{2}=($ $\varphi _{1}^{\dagger }\varphi
_{1}+\varphi _{2}^{\dagger }\varphi _{2}+\varphi _{3}^{\dagger }\varphi
_{3})^{2}.$ Corresponding to $\underline{1}^{\prime }\times \underline{1}%
^{\prime \prime },$ we have $(\varphi _{1}^{\dagger }\varphi _{1}+\omega
\varphi _{2}^{\dagger }\varphi _{2}+\omega ^{2}\varphi _{3}^{\dagger
}\varphi _{3})(\varphi _{1}^{\dagger }\varphi _{1}+\omega ^{2}\varphi
_{2}^{\dagger }\varphi _{2}+\omega \varphi _{3}^{\dagger }\varphi _{3})$
giving rise to $q$ and the quartic invariant $q^{\prime }=\varphi
_{1}^{\dagger }\varphi _{1}\varphi _{2}^{\dagger }\varphi _{2}+\varphi
_{2}^{\dagger }\varphi _{2}\varphi _{3}^{\dagger }\varphi _{3}+\varphi
_{3}^{\dagger }\varphi _{3}\varphi _{1}^{\dagger }\varphi _{1}.$ Next,
corresponding to $\underline{3}\times \underline{3}$ and $\underline{3}%
\times \underline{3}$ we have $q^{\prime \prime }=(\varphi _{1}^{\dagger
}\varphi _{2},\varphi _{2}^{\dagger }\varphi _{3},\varphi _{3}^{\dagger
}\varphi _{1}).(\varphi _{2}^{\dagger }\varphi _{1},\varphi _{3}^{\dagger
}\varphi _{2},\varphi _{1}^{\dagger }\varphi _{3})=|\varphi _{1}^{\dagger
}\varphi _{2}|^{2}+|\varphi _{2}^{\dagger }\varphi _{3}|^{2}+|\varphi
_{3}^{\dagger }\varphi _{1}|^{2}$ and $q^{\prime \prime \prime }=(\varphi
_{1}^{\dagger }\varphi _{2},\varphi _{2}^{\dagger }\varphi _{3},\varphi
_{3}^{\dagger }\varphi _{1}).(\varphi _{1}^{\dagger }\varphi _{2},\varphi
_{2}^{\dagger }\varphi _{3},\varphi _{3}^{\dagger }\varphi _{1})=(\varphi
_{1}^{\dagger }\varphi _{2})^{2}+(\varphi _{2}^{\dagger }\varphi
_{3})^{2}+(\varphi _{3}^{\dagger }\varphi _{1})^{2}.$ The 5th invariant is the complex conjugate of $q^{\prime \prime \prime }$.

Thus, the most general Higgs potential is given by $V=-\mu ^{2}s+\lambda
q+\lambda ^{\prime }q^{\prime }+\lambda ^{\prime \prime }q^{\prime \prime }+%
\frac{1}{2}(\lambda ^{\prime \prime \prime }q^{\prime \prime \prime }+h.c.)$
Assuming that the 3 $\varphi $'s all point in the same direction within $%
SU(2)$, then we have $V=-\mu ^{2}(v_{1}^{2}+v_{2}^{2}+v_{3}^{2})+\lambda
(v_{1}^{4}+v_{2}^{4}+v_{3}^{4})+\tilde{\lambda}%
(v_{1}^{2}v_{2}^{2}+v_{2}^{2}v_{3}^{2}+v_{3}^{2}v_{1}^{2})$ where $\tilde{%
\lambda}\equiv 2\lambda +\lambda ^{\prime }+\lambda ^{\prime \prime
}+\lambda ^{\prime \prime \prime }.$ For the sake of simplicity, we will take 
$\lambda ^{\prime \prime \prime }$ and the various $v$'s to be real, since
our focus here is not on $CP$ violation.

It is then straightforward though tedious to calculate the value of $V$ and
the eigenvalues $\Omega $ of the second derivative matrix $\frac{\partial
^{2}V}{\partial v_{\alpha }\partial v_{\beta }}$ evaluated at the three
mimina of interest: $E$ $:\{v_{1}=v_{2}=v_{3}=v\},$ $U:%
\{v_{1}=v,v_{2}=v_{3}=0\},$ and $P:\{v_{1}=v_{2}=v,v_{3}=0\}.$ We find

$E:v^{2}=\frac{\mu ^{2}}{2(\lambda +\tilde{\lambda})},V|_{E}=-\frac{3\mu ^{4}%
}{4(\lambda +\tilde{\lambda})},\Omega $ $=$ $[4\mu ^{2},\frac{2\mu
^{2}(2\lambda -\tilde{\lambda})}{\lambda +\tilde{\lambda}},\frac{2\mu
^{2}(2\lambda -\tilde{\lambda})}{\lambda +\tilde{\lambda}}]$

$U:$ $v^{2}=\frac{\mu ^{2}}{2\lambda },V|_{U}=-\frac{\mu ^{4}}{4\lambda },$ $%
\Omega $ $=$ $[4\mu ^{2},\frac{\mu ^{2}(\tilde{\lambda}-2\lambda )}{\lambda }%
,\frac{\mu ^{2}(\tilde{\lambda}-2\lambda )}{\lambda }]$ and

$P:v^{2}=\frac{\mu ^{2}}{2\lambda +\tilde{\lambda}},V|_{P}=-\frac{\mu ^{4}}{%
2\lambda +\tilde{\lambda}},$ $\Omega $ $=$ $[4\mu ^{2},\frac{2\mu ^{2}(%
\tilde{\lambda}-2\lambda )}{2\lambda +\tilde{\lambda}},\frac{2\mu ^{2}(%
\tilde{\lambda}-2\lambda )}{2\lambda +\tilde{\lambda}}]$

We note that by choosing $\tilde{\lambda}<0$ and sufficiently close to $%
-\lambda $ or by not doing this we could set $V|_{E}$ much lower than $%
V|_{U} $ or vice versa. On the other hand, for $P$ to be a minimum, we need $%
\tilde{ \lambda}-2\lambda >0$, which would make $V|_{U}$ lower than $V|_{P}.$
It appears that in this simple one Higgs doublet case, $P$ is never the true
minimum. Of course, in all the models we discussed, we have to introduce
more than one Higgs doublets and so presumably almost anything is possible
by coupling the various doublets together.

\ \bigskip

{\bf {\large Acknowledgments\medskip }}

I would like to thank Ray Volkas for inviting me to the Tropical Neutrino
Conference held in Palm Cove, Australia, for tirelessly urging me to look at
neutrino physics again, and for alerting me to interesting papers on the
subject as they appeared. I am grateful to Ernest Ma for a careful reading
of the manuscript and to Steve Hsu for a helpful discussion. Finally, I
thank Matt Pillsbury for some discussions about neutrino mixing. This work
was supported in part by the National Science Foundation under grants PHY
99-07949 and PHY00-98395.

\smallskip

{\bf {\large References}}


\begin{references}
\bibitem{4thpower}  Note that $A^{4}=9\omega I$ so that $A$ is up to an
overall factor the matrix 4th root of the identity.
\end{references}
\end{document}